\documentclass[journal=nalefd,manuscript=article]{achemso}

\usepackage{chemformula} 
\usepackage[T1]{fontenc} 
\usepackage{pdfpages}

\author{Aravind Raji}
\email{aravind.raji@universite-paris-saclay.fr}
\affiliation[1]
{Laboratoire de Physique des Solides, CNRS, Université Paris-Saclay, 91405 Orsay, France}
\alsoaffiliation[2]
{Synchrotron SOLEIL, L’Orme des Merisiers, BP 48 St Aubin, 91192 Gif sur Yvette, France}
\author{Guillaume Krieger}
\affiliation[3]
{Universite de Strasbourg, CNRS, IPCMS UMR 7504, F-67034 Strasbourg, France}
\author{Nathalie Viart}
\affiliation[3]
{Universite de Strasbourg, CNRS, IPCMS UMR 7504, F-67034 Strasbourg, France}
\author{Daniele Preziosi}
\email{daniele.preziosi@ipcms.unistra.fr}
\affiliation[3]
{Universite de Strasbourg, CNRS, IPCMS UMR 7504, F-67034 Strasbourg, France}
\author{Jean-Pascal Rueff}
\affiliation[2]
{Synchrotron SOLEIL, L’Orme des Merisiers, BP 48 St Aubin, 91192 Gif sur Yvette, France}
\alsoaffiliation[4]
{LCPMR, Sorbonne Université, CNRS, 75005 Paris, France}
\author{Alexandre Gloter}
\email{alexandre.gloter@universite-paris-saclay.fr}
\affiliation[1]
{Laboratoire de Physique des Solides, CNRS, Université Paris-Saclay, 91405 Orsay, France}

\title{Charge distribution across capped and uncapped infinite-layer \\neodymium nickelate thin films}

\abbreviations{STEM-EELS, HAADF, HAXPES}

\DeclareUnicodeCharacter{0301}{\'{e}}
\begin{document}

\begin{abstract}
Charge ordering (CO) phenomena have been widely debated in strongly-correlated electron systems mainly regarding their role in high-temperature superconductivity. Here, we elucidate the structural and charge distribution in NdNiO$_{2}$ thin films prepared with and without capping layers, and characterized by the absence and presence of CO. Our microstructural and spectroscopic analysis was done by scanning transmission electron microscopy-electron energy loss spectroscopy (STEM-EELS) and hard x-ray photoemission spectroscopy (HAXPES). Capped samples show Ni$^{1+}$, with an out-of-plane (o-o-p) lattice parameter of around 3.30 \AA \ indicating good stabilization of the infinite-layer structure. Bulk-sensitive HAXPES on Ni-2p shows weak satellite feature indicating large charge-transfer energy. The uncapped samples evidence an increase of the o-o-p parameter up to 3.65 \AA \ on the thin-film top, and spectroscopies show signatures of higher valence in this region (towards Ni$^{2+}$). Here, 4D-STEM demonstrates (3,0,3) oriented stripes which emerge from partially occupied apical oxygen. Those stripes form quasi-2D coherent domains viewed as rods in the reciprocal space with $\Delta\text{q}_{z} \approx 0.24$ r.l.u. extension located at \textbf{Q} = ($\pm \frac{1}{3},0,\pm \frac{1}{3}$) r.l.u. and \textbf{Q} = ($\pm \frac{2}{3},0,\pm \frac{2}{3}$) r.l.u. The stripes associated with oxygen re-intercalation concomitant with hole doping suggests a possible link to the previously reported CO in infinite-layer nickelate thin films.
\end{abstract}

\section{Introduction}
Discovery of superconductivity in Sr-doped infinite-layer nickelate thin films \cite{li2019superconductivity} has drawn an upsurge of interest being infinite-layer (IL) nickelates structurally and electronically analogous to cuprates. Apart from superconductivity, recent studies have also revealed the existence of charge ordering phenomena and magnetic excitations in IL-nickelate thin films \cite{krieger2022charge,tam2022charge,lu2021magnetic,rossi2022broken}, the amplitudes of which mostly depend upon the particular sample preparation, $i.e.$ presence/absence of a SrTiO$_{3}$ (STO) capping-layer and doping level. In particular, charge order (CO) has been observed in uncapped-IL NdNiO$_{2}$ samples and it is not much pronounced in the capped ones, which on the contrary host dispersing magnetic excitation with a bandwidth of circa 200 meV \cite{krieger2022charge,lu2021magnetic}. Such a marked dichotomy naturally raises the question of the structural stability of the uncapped IL-nickelate thin films, compared to that of the capped ones. Indeed, by considering previous reports about the perovskite nickelates \cite{wang2016oxygen,kotiuga2019carrier,shin2022magnetic}, one could infer on similar grounds, peculiar oxygen dynamics in IL-nickelates, such as vacancies or incomplete re-intercalation. The latter may distort the IL-crystalline structure, subsequently their electronic properties. Synthesis of IL-nickelates is obtained via a so-called topotactic reduction of the precursor perovskite RNiO$_{3}$ (RNO3, R being a rare-earth) thin films which involves removing apical oxygens in RNO3 thereby reducing it to the IL-nickelate RNiO$_{2}$ (IL-RNO2) \cite{li2019superconductivity}. Such a pathway involves reduction of the nominal valence of Ni from 3+ (in RNO3) to 1+ in (RNO2). This partially unstable state \cite{hayward1999sodium, subedi2023possible} can be prone to several structural and/or electronic reconstructions, and for a stable IL-RNO2 an SrTiO$_{3}$ (STO) capping-layer has been used in the 2-25 nm thickness range \cite{lee2020aspects, goodge2021doping}. On these grounds, the possible structural reconstructions in the uncapped IL-nickelate thin films stays unexplored. An atomically resolved structural and spectroscopic study which compares both capped and uncapped samples is necessary to obtain new insights to delineate a better understanding of such aforementioned contrasting differences.  

Here we have investigated STO-capped and uncapped IL-NdNiO$_{2}$ thin films grown onto STO single crsytals as substrate, and hereafter referred to as c-NNO2 and uc-NNO2, respectively. By combining spectroscopic and microscopic techniques, specifically hard x-ray photoemission spectroscopy (HAXPES), scanning transmission electron microscopy (STEM) with monochromated electron energy loss spectroscopy (EELS), we conclude that the measured structural and electronic modifications are linked to a specific O re-intercalation path and charge reconstruction, which could contribute to the observed CO.
\section{Results and discussion}
\begin{figure}[ht]
\centering
\includegraphics[width=\textwidth]{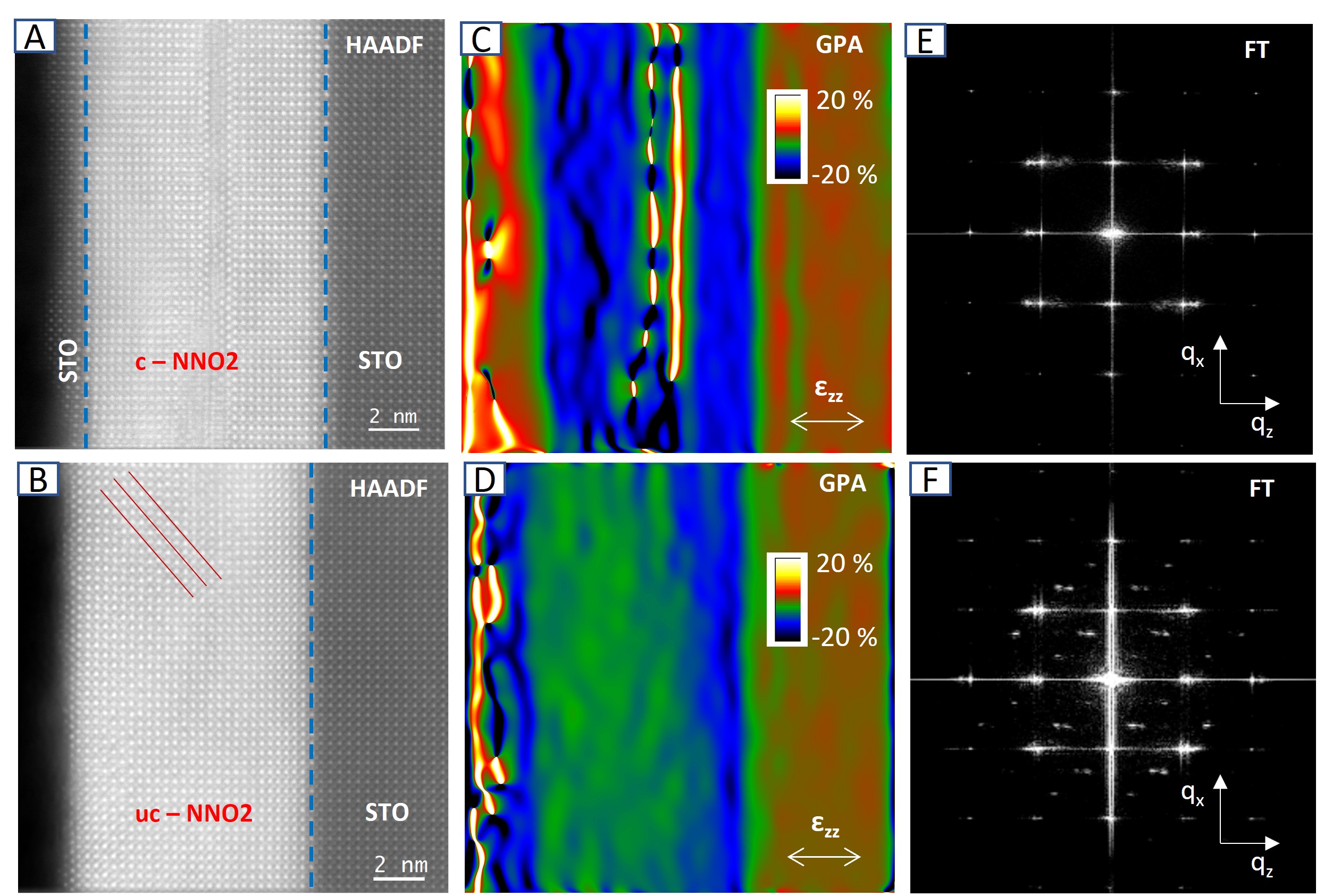}
\caption{Comparison of capped (c-NNO2) and uncapped (uc-NNO2) infinite-layer nickelate thin films by HAADF, GPA and FT. (A and B) HAADF images showing the (A) c-NNO2 and (B) uc-NNO2, the faint contrast (3 0 3) stripe pattern as indicated by the red lines can be seen here. (C and D) Maps of the o-o-p strain by GPA in (C) c-NNO2 and (D) uc-NNO2. Both capped and uncapped samples show a higher compression close to the film-substrate interface. (C) c-NNO2 shows uniform compression of about 16\% all over the sample, except for the small defect zone in the middle (D) uc-NNO2 shows a less compressed zone towards the surface from the interface. (E and F) Fourier Transform of the HAADF image of the (E) c-NNO2 and (F) uc-NNO2.}
\label{fig:Figure 1}
\end{figure}
We start by discussing the real space structural aspects of both c-NNO2 and uc-NNO2 samples as obtained from STEM-HAADF imaging and geometrical phase analysis (GPA) \cite{hytch1998quantitative}. As shown in the STEM-HAADF images in Figs.\ref{fig:Figure 1}A and B, clear structural differences are observed between these two samples. The c-NNO2 shows an IL-structure in the majority of the thin-film, except the small off tilt-like defect in the middle as reported \cite{zeng2020phase}. On the other hand, the uc-NNO2 shows an IL-structure near the interface, with certain periodic faint contrast stripes on the top part of the thin-film which will be discussed in more details in the following section. Fig.\ref{fig:Figure 1}C shows the o-o-p strain maps ($\epsilon_{zz}$) obtained by GPA in the c-NNO2, with reference to the STO o-o-p parameter of 3.91 \AA. \ A compression of around 16\% giving an o-o-p parameter of 3.30 \AA \ is observed throughout the thin-film (blue region in Fig.\ref{fig:Figure 1}C), except in the defect region in the middle, indicating overall a good infinite-layer crystalline quality of this c-NNO2. Fig.\ref{fig:Figure 1}D shows such a strain map of the uc-NNO2, where a compression of around 16\% is observed near the interface, giving an infinite-layer o-o-p parameter of 3.30 \AA.\ However, the map shows a decrease of the compression to around 7\% giving an o-o-p parameter of 3.65 \AA \ on the top (green region in Fig.\ref{fig:Figure 1}D). The region of this o-o-p expansion coincides with the region where we observe certain low contrast stripes for this uc-NNO2 sample. The HAADF image in Fig.\ref{fig:Figure 1}B for the uc-NNO2 is from a region with more extended stripes, while other regions show stripes more restrained extending of only ca. 2 nm from the surface (Supporting Information, Figure S1). The two samples show no differences in the in-plane strain ($\epsilon_{xx}$) map (Supporting Information, Figure S2), and a homogeneous in-plane parameter of 3.91 \AA \ matching with STO is observed throughout the thin-film. The Fourier transform analysis shown in Figs.\ref{fig:Figure 1}E and F for the c-NNO2 and uc-NNO2, respectively further describes the differences observed in HAADF and GPA. The c-NNO2 shows a pair of spots along q$_{z}$ around (1 0 1) r.l.u. that matches with the different o-o-p parameters of STO and the infinite-layer structure. In contrast, the uc-NNO2 shows three spots along q$_{z}$ around (1 0 1) r.l.u. indicating the o-o-p parameters of 3.91 \AA, \ 3.30 \AA, \ and 3.65 \AA \ respectively from the STO, infinite-layer and the stripe region. Here, we also observe extended spots located at q$_{x}$ = q$_{z}$ =  $\frac{1}{3}$ r.l.u., and q$_{x}$ = q$_{z}$ = $\frac{2}{3}$ r.l.u. The reduced lattice units (r.l.u.), is defined with in plane components a = b = 3.91 \AA, \ and the out-of-plane (o-o-p) lattice constant c = 3.65 \AA. \ We chose the o-o-p as 3.65 \AA \  because this is the parameter of the stripe region. Interestingly, similar faint contrast stripes can be observed in the STEM-HAADF image of other uncapped nickelate samples where CO is reported (SI Fig. S5 of \cite{tam2022charge}), but were unnoticed. It further anticipates the connection between CO and these (3 0 3) stripes.   

As mentioned by Krieger $et$ $al.$ in \cite{krieger2022charge}, the CO signal decreases with the level of Sr-doping on NdNiO$_{2}$. In (Supporting Information, Figure S4), we show our STEM-HAADF and GPA analysis on a 5\% Sr-doped uncapped Nd$_{0.95}$Sr$_{0.05}$NiO$_{2}$ (uc-NSNO2) sample. We don't observe any stripes or o-o-p expansion as in the case of uc-NNO2. There are certain fluorite and Ruddlesden-Popper defects on the top of the sample, which are reportedly observed in these systems \cite{lee2020aspects}. This also rises possible connection of CO to the presence of such charge-stripes as observed in uc-NNO2. Another aspect here is about the structural stability of undoped samples compared to the doped ones. As mentioned in \cite{tam2022charge}, a comparative study between uncapped samples that are both doped and undoped, indicates a difference in the o-o-p parameter for certain undoped samples. Also, as shown in the (Supporting Information, Figure S3 and Figure S4), the Fourier transform shows no stripe intensities in the regions dominated by additional Fluorite-type defects, suggesting that defects can reduce re-intercalation processes. In the case of 5\% Sr-doped sample, the presence of the dopant probably changes the chemical pathway for O re-intercalation. Doping possibly causes the stabilization of particular phases, that prevent the formation of the stripe order.     
\begin{figure}[p]
\centering
\includegraphics[width=\textwidth]{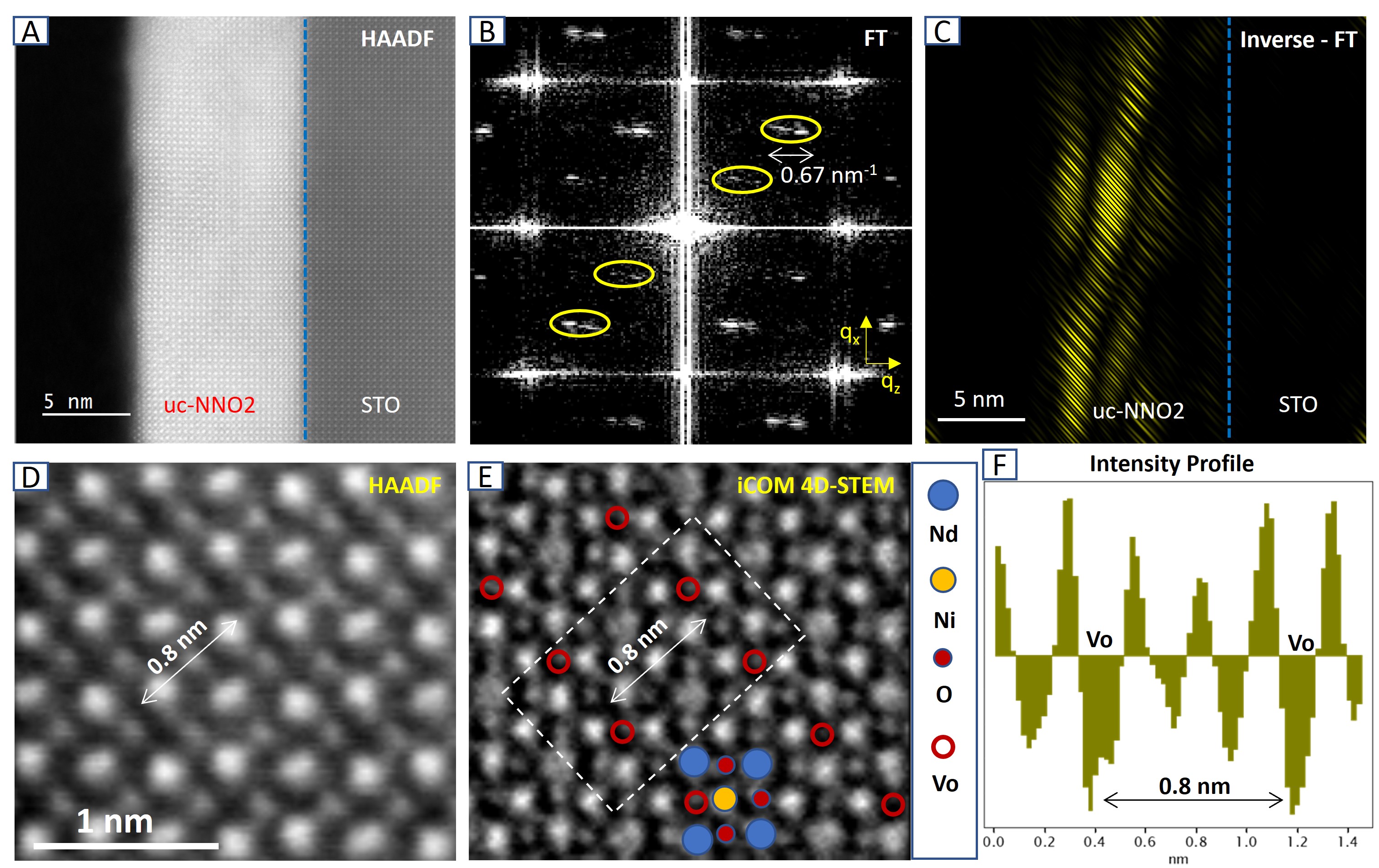}
\caption{Combined reciprocal space and 4D-STEM analysis of the stripe region of the uc-NNO2. (A) HAADF image showing the uc-NNO2 with stripes on the top part of the thin film. (B) Magnified Fourier Transform image showing an expanded unit cell and periodic rods located at q$_{x}$ = q$_{z}$ =  $\frac{1}{3}$ r.l.u., and q$_{x}$ = q$_{z}$ = $\frac{2}{3}$ r.l.u., with a $\Delta\text{q}_{z}$ = 0.24 r.l.u., (C) The Inverse-Fourier Transform of the rods showing coherent domains that extends over 10 x 1.5 nm in size. (D) A high magnification HAADF image from the stripe region (highlighted in A)  showing periodic low contrast that gives rise to the stripes. (E) 4D-STEM integrated Centre of Mass (iCOM) image in the same region showing partial apical oxygen intercalation, with absence of oxygen at the apical sites along the stripes. (F) An intensity profile of the highlighted region of interest in the iCOM image that shows enhanced negative intensity further confirming oxygen vacancies along the stripes.}
\label{fig:Figure 2}
\end{figure} 
Fig.\ref{fig:Figure 2}B shows a magnified Fourier transform analysis of the HAADF image showing stripes in the uc-NNO2 (Fig.\ref{fig:Figure 2}A). Here, the observed periodicity of the stripes contributes to extended rods in reciprocal space located at q$_{x}$ = q$_{z}$ =  $\frac{1}{3}$ r.l.u., and q$_{x}$ = q$_{z}$ = $\frac{2}{3}$ r.l.u., with a $\Delta\text{q}_{z}$ $\approx$ 0.24 r.l.u. In the inverse-Fourier transform analysis Fig.\ref{fig:Figure 2}C, this appears as coherent domains that have a typical area of 10 x 1.5 nm$^2$. From our observations throughout the thin-film, these coherent domains appear as quasi-2D sheets of ca. 1.5 - 2 nm extension along the o-o-p direction. When the stripes areas extend over a substantial part of the thin film, several sheets are coexisting within the depth as shown in Fig.\ref{fig:Figure 2}C, but areas with only one sheet can be also observed (Supporting Information, Figure S1). 
The stripe wavevector values reported here are in match with the CO wave-vector of \textbf{Q} = ($\pm \frac{1}{3}$, 0) r.l.u. as found in the previous studies of these samples \cite{krieger2022charge}, and others \cite{tam2022charge, rossi2022broken} by studying the quasi-elastic scattering intensity of resonant inelastic x-ray scattering (RIXS) experiments. Recently, a q$_{z}$ resolved resonant x-ray scattering (RXS) study has been carried out on an infinite-layer PrNiO$_{2}$ thin film in which the CO has the \textbf{Q} = ($\pm \frac{1}{3}$, 0, 0.365) r.l.u. wavevector \cite{ren2023symmetry}. The reported value of q$_{z}$ = 0.365 r.l.u. is within the extension of the rod in Fig.\ref{fig:Figure 2}B, extending between q$_{z}$ = 0.22 to 0.46 r.l.u. In the Fourier transform in Fig.\ref{fig:Figure 1}E, no such rods are observed in the reciprocal space of the c-NNO2 sample. This goes in hand with the absence of CO reported in these c-NNO2 samples \cite{krieger2022charge}. Fig.\ref{fig:Figure 2}D shows a magnified HAADF image in the stripe region, showing a faint dark contrast along the stripes, and between the rare-earth sites. A better understanding of this region is obtained from high-resolution 4D-STEM analysis. By collecting the whole diffraction pattern at each pixel, one can obtain a high resolution real space atomic mapping with good oxygen contrast \cite{jia2023real,yang2021determination}. Here, in Fig.\ref{fig:Figure 2}E, by employing such an integrated Centre of Mass (iCOM) analysis by 4D-STEM, we have a clear atomic-level mapping including oxygen. It indicates periodic partial intercalation of apical oxygen in the IL crystalline structure. If we consider a perovskite NNO3 structure, this could be interpreted as periodic apical oxygen vacancies (Vo). This finding is of paramount importance as it has been well demonstrated the influence of Vo in changing the local electronic structure especially in perovskite nickelates \cite{kotiuga2019carrier}. Since these Vo run along the stripes, they have the same periodicity as them, contributing to intensities at \textbf{Q} = ($\pm \frac{1}{3}$ 0 $\pm \frac{1}{3}$) r.l.u. and \textbf{Q} = ($\pm \frac{2}{3}$ 0 $\pm \frac{2}{3}$) r.l.u. in the reciprocal space.

The oxygen re-intercalations observed on the stripe-region of uc-NNO2, probably induce similar hole doping effects as in \cite{zhao2019charge} changing the Ni valence at these sites. The charge distribution is reflected in the spectroscopic studies, XAS showing a mixed valence and RIXS experiments already demonstrating a broken symmetry of the uc-NNO2 \cite{krieger2022charge} in reciprocal space. Since, now we have an understanding of the real space location of these stripes, HAXPES and monochromated STEM-EELS can probe the concomitant charge modulation as well.

\begin{figure}[ht]
\centering
\includegraphics[width=\textwidth]{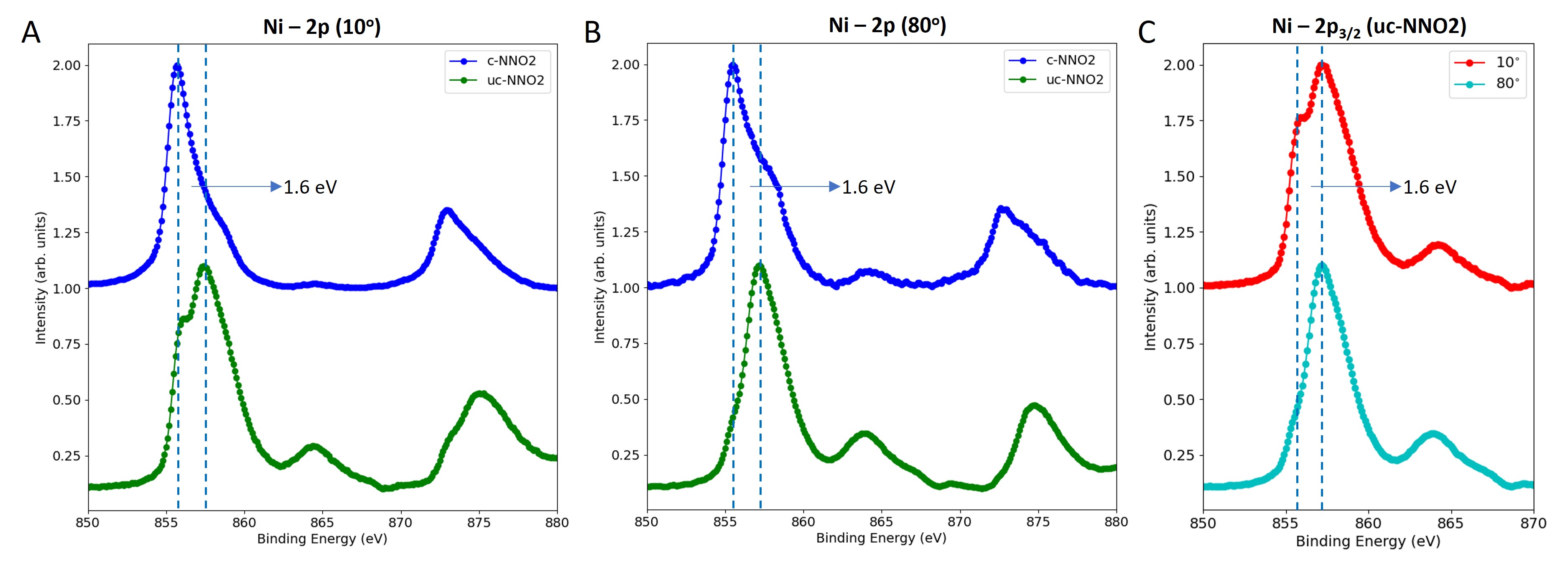}
\caption{Hard X-ray Photoemission (HAXPES) comparison of the Ni 2p core level of the c-NNO2 and uc-NNO2 samples. (A and B) Comparison of Ni 2p photoemission in (A) bulk sensitive 10$^{\circ}$ incidence condition and (B) surface sensitive 80$^{\circ}$ incidence condition from c-NNO2 and uc-NNO2 samples. (C) A magnified comparison of Ni 2p$_{3/2}$ photoemission at these two incidence conditions in the uc-NNO2 sample, clearly showing the potential decrease of the low binding energy shoulder in the 80$^{\circ}$ condition, indicating higher valence at the surface of uc-NNO2.}
\label{fig:Figure 3}
\end{figure}

HAXPES gives a depth resolved macroscopic understanding of the influence of this structural modification on the local electronic structure. Figs.\ref{fig:Figure 3}A and B shows the HAXPES data in both bulk-sensitive and surface-sensitive configurations. This is done to better discriminate signals stemming from the top and the bulk/interface regions of the samples. The bulk sensitive mode is obtained at 10$^{\circ}$ incidence angle of the incoming photons, with an estimated probing depth of 10 nm at 3000 eV as obtained by SESSA simulations \cite{smekal2005simulation}. In the surface sensitive mode, the incidence angle of 80$^{\circ}$ gives a probing depth of circa 2 nm at the same photon energy to get similar energy resolution. In both conditions, we compare the Ni 2p core level for the c-NNO2 and uc-NNO2. As expected, there are strong differences between the two samples and incidence conditions of the photons. 

In the bulk-mode, it can be seen that the main Ni 2p$_{3/2}$ peak shows a shift between capped and uncapped, with a shoulder at a high binding energy (HBE) and low binding energy (LBE) respectively. The separation between them is almost 1.6 eV. This difference gets stronger as we go to surface-sensitive configuration, indicating a strong difference between the top part of the uc-NNO2 with that of the c-NNO2. The shift to high binding energy in the uc-NNO2 indicates a more oxidized Ni species, closer to Ni$^{2+}$ than to Ni$^{1+}$, as in the case of IL-NNO2 structure. In Fig.\ref{fig:Figure 3}C, this difference is depicted more clearly by comparing the uc-NNO2 Ni-2p spectra at 10$^{\circ}$ and 80$^{\circ}$. The decrease of LBE shoulder as we go more surface indicates the presence of more Ni$^{2+}$ or higher valence at the top of the thin-film. Such a strong trend is not observed in the c-NNO2. A similar comparison of the Nd 3d photoemission spectra for both samples rendered no such peculiar differences of the core level (Supporting Information, Figure S5).

The c-NNO2 being a perfect infinite-layer thin film, hosts Ni$^{1+}$ in a 3d$^{9}$ configuration. It is to note that the c-NNO2 exhibit a very weak satellite peak located at almost 9 eV  at higher binding energy from the main peak. It is rather different from previously reported soft X-ray PES that was having a strong satellite at only ca. 6 eV from the main edge \cite{higashi2021core}. 
The discrepancy might occur, since our hard X-ray photoemission probes the bulk part of the capped sample, which is structurally a perfect infinite-layer. A weaker satellite peak at higher energy would indicate larger charge-transfer energy, pleading for a stronger Mott-Hubbard character\cite{zaanen1985band} of the NNO2 than the one already infered from the PES \cite{higashi2021core}, but more in accordance with the previously XAS and RIXS report of holed-doped system\cite{rossi2021orbital}.

In the case of uc-NNO2, as observed by 4D-STEM, there is still a partial presence of apical oxygen, resulting in hole doping that is changing the valence towards Ni$^{2+}$. The HAXPES of Fig.\ref{fig:Figure 3} is characterized by a strong satellite peak located at a higher binding energy around 865 eV. The Ni species may be described as a mixture of Ni d$^{8}$ and Ni d$^{9}\underline{L}$ configuration, where $\underline{L}$ describes a hole in the ligand orbital, resulting in satellite and main edges. Comparing the Ni 2p photoemission spectrum from the top part of the uc-NNO2 with previously reported  cases for NiO \cite{van1986comparison} and reduced SrNiO$_{3}$ nickelate \cite{wang2021spontaneous}, the present case is more comparable to the latter. 
It indicates that the Ni ion here is not exactly as for Ni$^{2+}$ in NiO which is characterized by a strong d$^{8}$ \cite{van1986comparison} ground state, but bear more ligand-hole contribution. HAXPES is highly sensitive to changes in local chemistry, but it also gives an averaged macroscopic signal. The strong LBE shoulder that appears together with the main peak in the uc-NNO2 at 10$^{\circ}$ incidence (bulk-mode) is indeed due to this averaging of signals from Ni sites in the IL-NNO2 near the interface and the higher valence Ni at the surface as due to possible oxygen re-intercalation. 
\begin{figure}[p]
\centering
\includegraphics[width=\textwidth]{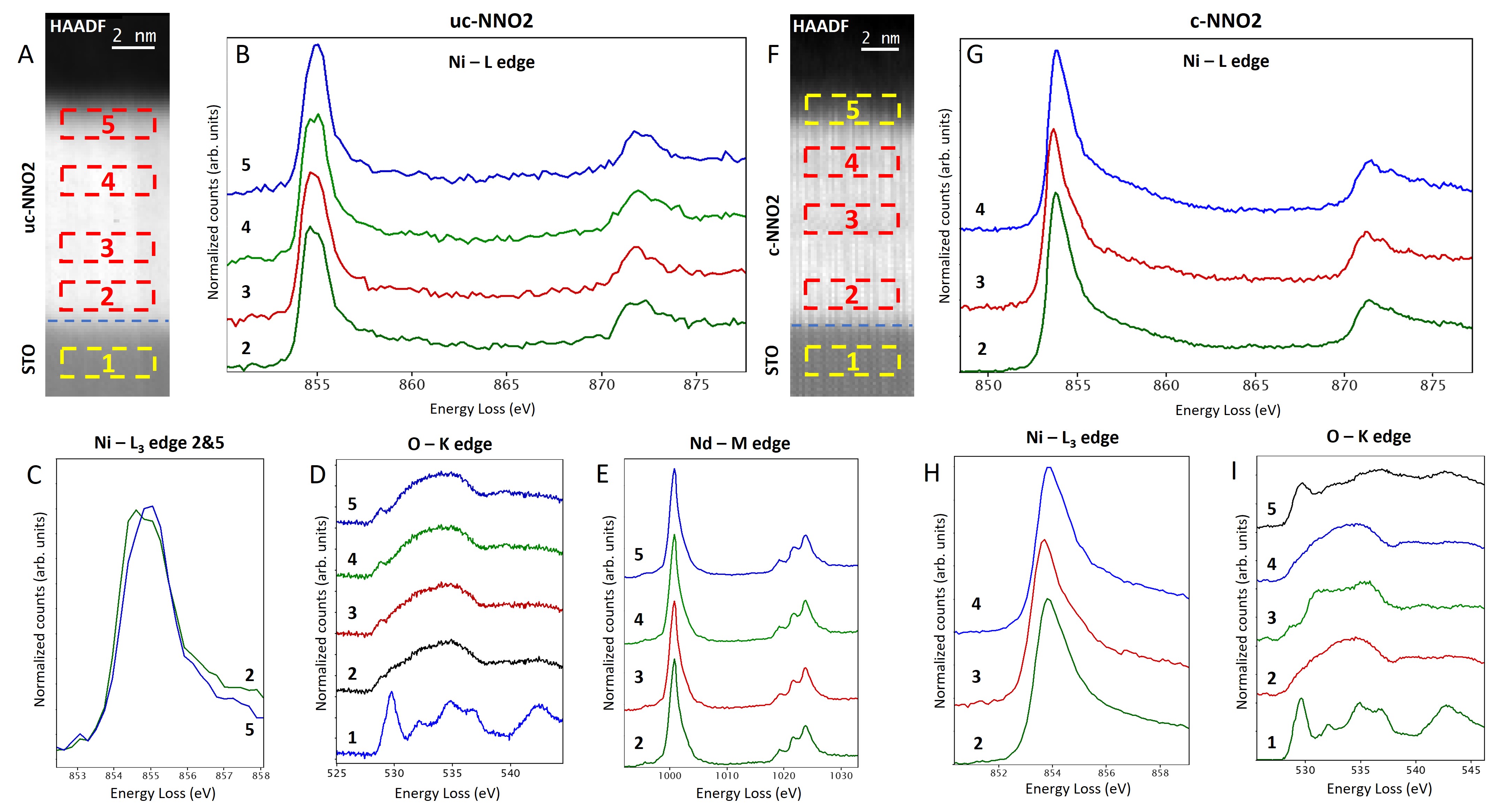}
\caption{Monochromated electron energy loss spectroscopy (EELS) fine structure analysis of uc-NNO2 and c-NNO2 samples. (A) HAADF image of the uc-NNO2 sample. (B-G) EELS Fine structure analysis of different regions labelled in the HAADF image of the uc-NNO2 sample. (B and C) Ni-L edge comaparison of the regions in the thin-film, (C) showing significant fine structure difference at the Ni-L$_{3}$ edge between the very interface region and the top of the thin-film. (D) O-K edge fine structure showing emergence of a strong pre-peak around the top part of the thin-film. (E) Nd-M edge fine structure showing no visible differences across the thin-film. (F) HAADF image of the c-NNO2 sample. (G-I) EELS Fine structure analysis of different regions labelled in the HAADF image of the c-NNO2 sample. (G and H) Ni-L edge comaparison of the regions in the thin-film, showing no significant differences between them. (I) O-K edge fine structure showing no differences at the two interfaces of the thin film, where absence of pre-peak in the fine structure indicates the existence of the perfect infinite-layer. A difference in fine structure is observed for the defect region in the middle and the spectrum looks similar as reported in \cite{goodge2021doping}.}
\label{fig:Figure 4}
\end{figure}
A higher spatial and energy resolved spectroscopic comparison between c-NNO2 and uc-NNO2 samples is obtained by monochromated STEM-EELS. In Fig.\ref{fig:Figure 4}, such a spatial variation at Ni-L, O-K, and Nd-M edges is compared between both samples. In Figs.\ref{fig:Figure 4}B and C, a clear difference has been observed for the Ni-L$_{3}$ edge in the uc-NNO2, as we compare the spectrum from near the substrate interface, and to the top part of the thin film. The peak maxima shows a shift to a higher energy loss at the top of the thin film. Fig.\ref{fig:Figure 4}D shows the O-K edge in this sample, where no pre-peak feature is observed near the interface as expected for an IL-NNO2, while having a stronger pre-peak feature around 527 eV going to the top of the thin film. In the case of the c-NNO2 sample, as shown in Figs.\ref{fig:Figure 4}G - I, we observe no differences for the Ni-L and O-K edges (except near the small defect area in the middle), confirming existing data\cite{goodge2021doping}. Interestingly, the O-K edge in the whole volume of the c-NNO2 sample shows no pre-peak feature except the central defective region, indicating the good stabilization of the IL crystalline structure.

The observed spatial spectroscopic differences in the uc-NNO2 are possibly caused by local electronic reconstructions due to oxygen intercalation into the infinite-layer structure. The shift of Ni L$_{3}$ edge to a higher energy loss indicates a hole-doping effect towards a dominant Ni$^{2+}$.

The pre-peak in O-K edge at 527 eV stems from the hybridization between the O 2p and Ni 3d orbitals \cite{de1989oxygen}. It is very strong in the case of perovskite NdNiO$_{3}$ with Ni being in a 3d$^{8}\underline{L}$ ground state in the metallic phase. It almost disappears for an ideal infinite-layer NdNiO$_{2}$, where the Ni is in a 3d$^{9}$ state, and also for NiO where the Ni is in a 3d$^{8}$ state\cite{van1986comparison}. The observation of a small pre-peak in O-K at the top part of uc-NNO2, while corresponding primary to a Ni$^{2+}$, can be explained as these Ni$^{2+}$ being a mixture of 3d$^{8}$ and also some 3d$^{9}\underline{L}$ electronic configuration, that is in accordance with the Ni 2p photoemission features in Fig.\ref{fig:Figure 3}. Similar to the Nd-3d photoemission spectra (Supporting Information, Figure S5), the Nd-M edge in Fig.\ref{fig:Figure 4}E shows no difference spatially in the uc-NNO2, and it can be attributed to the low sensitivity of rare earth elements to such local electronic/structural reconstructions.
\subsection{Outline}
Our multidimensional analysis demonstrates charge disproportionation associated with the formation of periodic stripes on the top region of an uc-NNO2 thin-film. The stripes are originating from partial oxygen re-intercalation into the IL-structure, hence, mimicking hole-doping effects. Spectroscopic signatures evidence the top part of the uc-NNO2 is mixed valence composed of Ni$^{1+}$ (d$^{9}$) and Ni$^{2+}$ (d$^{8}$ with possibly some d$^{9}\underline L$ configuration due to a larger degree of  hybridization). Since a possible origin of this additional charge on Ni may be attributed to oxygen re-intercalation (hole-doping), this charge distribution is influenced by the (3 0 3) periodicity of the stripes. The stripes form quasi-2D coherent domains of diverse spatial extensions throughout the thin-film. They give extended rods in reciprocal space at \textbf{Q} = ($\pm \frac{1}{3}$ 0 $\pm \frac{1}{3}$) r.l.u. and \textbf{Q} = ($\pm \frac{2}{3}$ 0 $\pm \frac{2}{3}$) r.l.u., with the former value fully comparable with the CO wavevector \textbf{Q} = ($\frac{1}{3}$, 0) r.l.u. observed in this sample. No such chemical and structural modification is observed in c-NNO2 and 5 \% Sr-doped uc-NSNO2, the latter closely connected with the observation of diminished CO. In the case of the uc-NNO2 studied here, the intercalation of oxygen to the apical sites essentially changes the local structure of these regions of the thin-film. Understanding the spatial configuration of oxygen intercalation, its energetic favourability and dynamics is another aspect along this direction.

\section{Experimental}
\subsection{Sample preparation}
Perovskite precursor thin films have been grown by Pulsed Laser Deposition technique assisted by Reflection High Energy Electron Diffraction onto STO substrates (CODEX) that underwent the standard HF-etching and annealing processes to obtain single terminated TiO$_{2}$-terrace morphology. The STO capping layer of three unit cells, when present, has been grown just after the perovskite nickelate and prior to the topotactic reduction as already mentioned elsewhere \cite{krieger2022synthesis}.
\subsection{High resolution STEM-EELS and 4D-STEM}
Cross sectional transition electron microscopy (TEM) lamellae were prepared using a focused ion beam (FIB) technique (D. Troadec at IEMN facility Lille, France; and at C2N, University of Paris-Saclay, France). Before FIB lamellae preparation, around 20 nm of amorphous carbon was deposited on top for protection. The HAADF imaging and 4D-STEM was carried out in a NION UltraSTEM 200 C3/C5-corrected scanning transmission electron microscope (STEM). The experiments were done at 100 keV with a probe current of approximately 10 pA and convergence semi-angles of 30 mrad. A MerlinEM (Quantum Detectors Ltd) in a 4×1 configuration (1024 × 256) has been installed on a Gatan ENFINA spectrometer mounted on the microscope \cite{tence2020electron}. The EELS spectra are obtained using the full 4 × 1 configuration and the 4D-STEM by selecting only one of the chips (256 × 256 pixels). For 4D-STEM, the EELS spectrometer is set into non-energy dispersive trajectories and we have used 6-bit detector mode that gives a diffraction pattern with a good signal to noise ratio without compromising much on the scanning speed. 
The monochromated EELS have been done using a NION CHROMATEM STEM at 100 keV with a probe current of approximately 30 pA, a convergence semi-angles of 25 mrad and an energy resolution around 70 meV. The EELS detection was also done with a MerlinEM in a 4×1 configuration (1024 × 256) that has been installed on a Nion IRIS spectrometer mounted on the microscope.
\subsection{HAXPES measurements}
The measurements were carried out at the GALAXIES beamline at the SOLEIL synchrotron \cite{rueff2015galaxies} on the HAXPES endstation \cite{ceolin2013hard} using a photon energy of 3000 eV, with an incidence angle of 10$^\circ$ for the bulk sensitive measurements and 80$^\circ$ for the surface sensitive measurements. The bulk sensitivity is defined from the SESSA simulations \cite{smekal2005simulation}, that gives a probing depth of around 10 nm for 10$^\circ$ incidence and around 2 nm for 80$^\circ$ incidence. About 95\% of the detected signal will be from the elements within these estimated probing depth. The synchrotron operated with a ring current of 450 mA, giving an intensity of 3.4 × 10$^{13}$ photons/s at 3000 eV, which was then reduced using a built-in filter to 5\% of the original intensity. The photoelectrons were detected using a SCIENTA Omicron EW4000 HAXPES hemispherical analyzer, and a Shirley background \cite{shirley1972high} was removed prior to fitting the core levels spectra.

\begin{acknowledgement}
This work was supported by the French National Research
Agency (ANR) through the ANR-21-CE08-0021-01 ‘ANR
FOXIES’ and, within the Interdisciplinary Thematic Institute
QMat, as part of the ITI 2021 2028 program of the University
of Strasbourg, CNRS and Inserm, it was supported by
IdEx Unistra (ANR 10 IDEX 0002), and by SFRI STRAT’US
project (ANR 20 SFRI 0012) and ANR-11-LABX-0058-NIE
and ANR-17-EURE-0024 under the framework of the French
Investments for the Future Program. A.R. acknowledges financing from LABEX NanoSaclay and H2020 for the doctoral funding. Nion CHROMATEM at LPS Orsay and the FIB at C2N, University of Paris-Saclay was accessed in the TEMPOS project framework (ANR 10-EQPX-0050). We acknowledge SOLEIL Synchrotron for provision of beamtime under proposals 20211467 and 20221574. We thank M. Salluzzo and G. Ghiringhelli for critical reading of the manuscript.

\end{acknowledgement}


\bibliography{references}



\section*{Supplementary material}

# Supporting Information

**Charge distribution across capped and uncapped infinite-layer neodymium nickelate thin films**

Aravind Raji[1,2,*], Guillaume Krieger[3], Nathalie Viart[3], Daniele Preziosi[3,*], Jean-Pascal Rueff[2,4], Alexandre Gloter[1,*]

1 Laboratoire de Physique des Solides, CNRS, Universite Paris-Saclay, 91405 Orsay, France
2 Synchrotron SOLEIL, L'Orme des Merisiers, BP 48 St Aubin, 91192 Gif sur Yvette, France
3 Universite de Strasbourg, CNRS, IPCMS UMR 7504, F-67034 Strasbourg, France
4 LCPMR, Sorbonne Universite, CNRS, 75005 Paris, France

*Corresponding authors (E-mail): aravind.raji@universite-paris-saclay.fr, daniele.preziosi@ipcms.unistra.fr, alexandre.gloter@universite-paris-saclay.fr

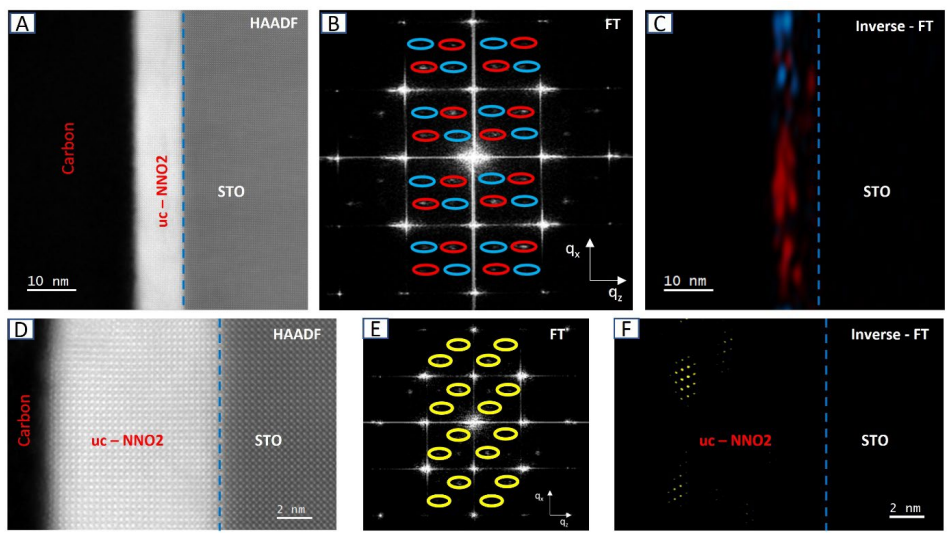

**Figure S1.** Distribution of different stripe areas in the thin-film. (A) Low magnification HAADF image of uc-NNO2 (B) Corresponding Fourier transform of the HAADF image. (C) Inverse - FT of the highlighted spots in the Fourier Transform showing differential distribution of stripe areas. The blue area corresponds to stripes oriented along (-3 0 3) planes while the red area are for (3 0 3) planes. (D) High magnification HAADF image showing a region with less stripes. (E) Corresponding Fourier transform of the HAADF image (F) Inverse - FT of the highlighted spots in the Fourier Transform showing quasi-2D stripe sheets near the surface that extends approximately 2 nm in z. Overall, The stripe areas have faint visibility in HAADF image (explaining why they have been unnoticed in the previous report of similar uncapped sample [1]), but are observed throughout the sample near the top surface. They have typically coherent extension over 1.5-2 nm along the z, resulting in rods with Δqz extension of ca. 0.24 r.l.u. Their observed in-plane extension goes from 5 nm to a few tens of nanometers resulting in quasi-2D sheets geometry.

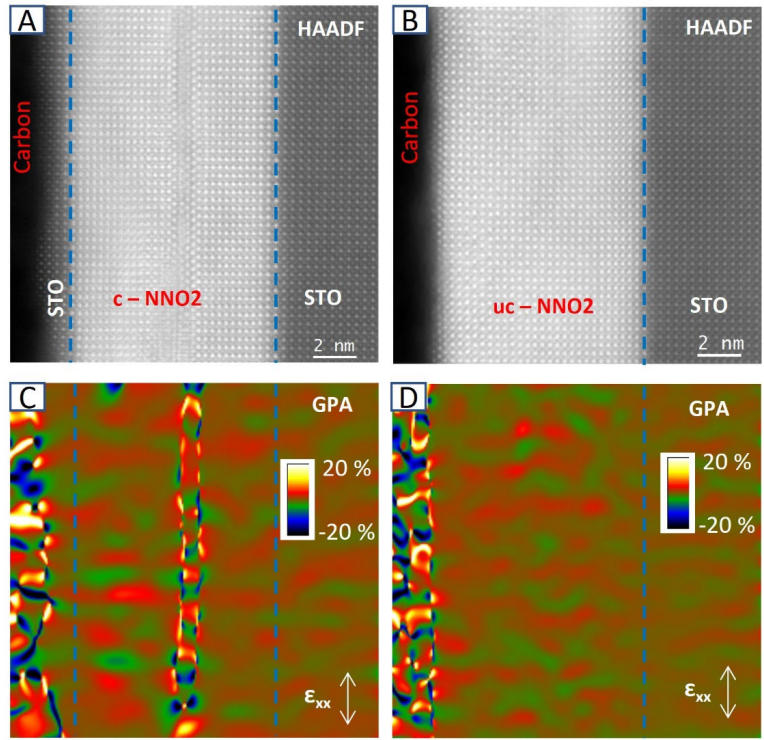

**Figure S2.** Comparison of in-plane strain of capped (c-NNO2) and uncapped (uc-NNO2) infinite-layer nickelate thin films by HAADF and GPA. (A) HAADF image of the uc-NNO2, (B) The in-plane strain map showing no spatial variations. (C) HAADF image of the c-NNO2, (D) The in-plane strain map showing only a small variation for the defect in the middle of thin film.

3
22

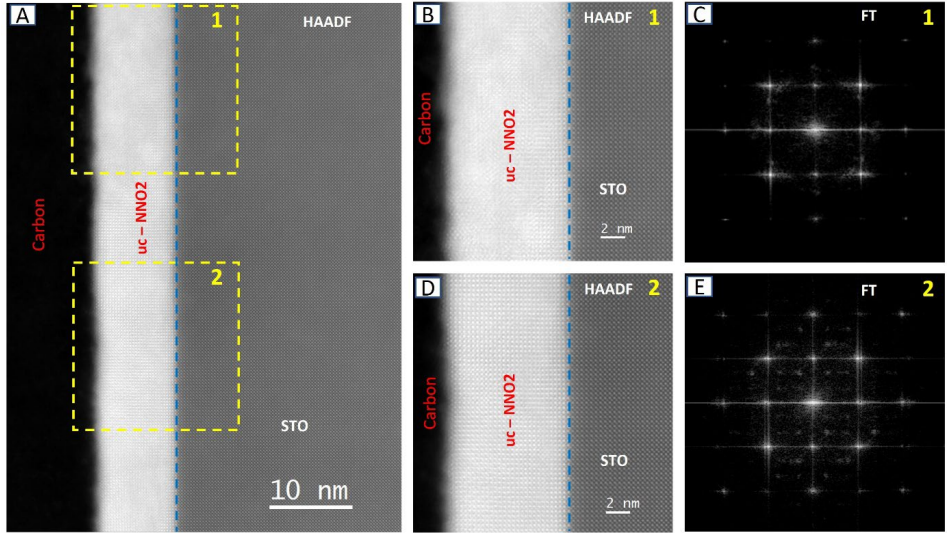

**Figure S3.** Comparison of Fourier Transform in a defective area with the stripe region of the thin-film. (A) A low magnification HAADF image of the uc-NNO2, (B) The magnified HAADF image of the region 1 as highlighted in (A), showing strong defects. (C) The corresponding Fourier transform of region 1, that does not show any intensity at Q = ($\frac{1}{3}$, 0, $\frac{1}{3}$) r.l.u. or Q = ($\frac{2}{3}$, 0, $\frac{2}{3}$) r.l.u. (D) The magnified HAADF image of the region 2 as highlighted in (A), showing stripes. (E) The corresponding Fourier transform of region 2, that shows intensities at Q = ($\frac{1}{3}$, 0, $\frac{1}{3}$) r.l.u. and Q = ($\frac{2}{3}$, 0, $\frac{2}{3}$) r.l.u.

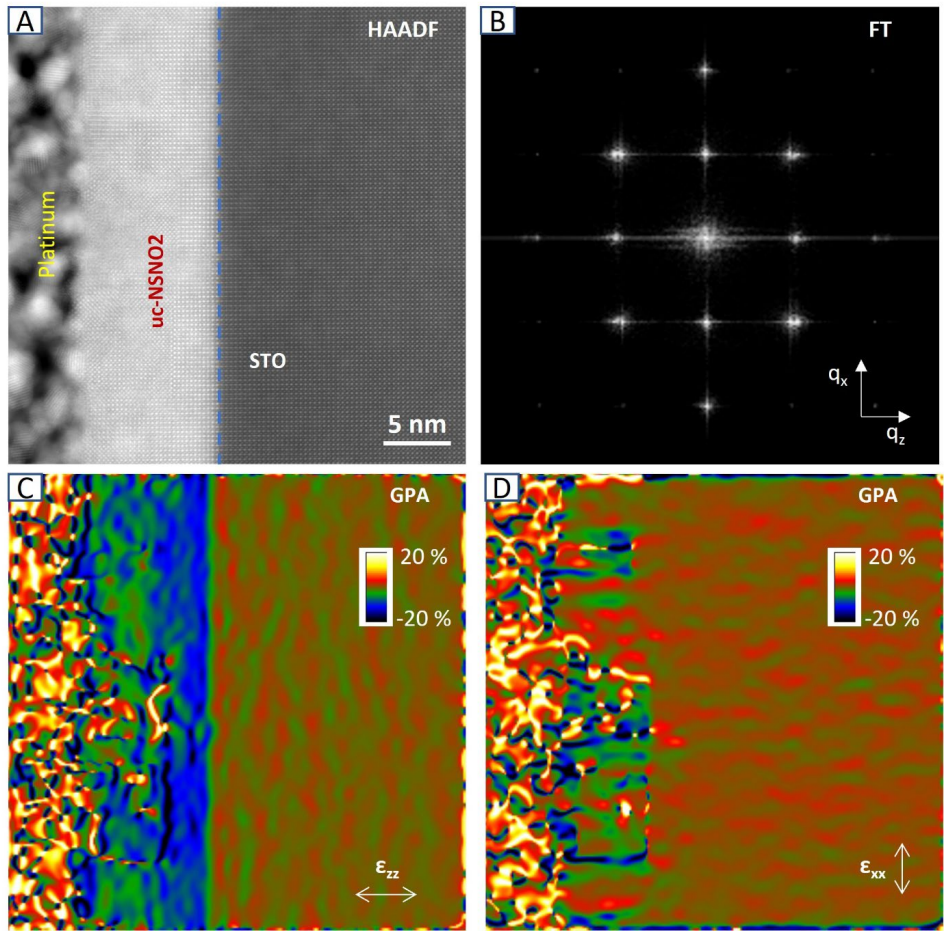

**Figure S4.** Comparison of in-plane and o-o-p strain of a 5% Sr-doped uncapped infinite-layer sample. (A) HAADF image showing certain defects and no stripe order on the top of the thin-film, (B) The corresponding Fourier Transform of the HAADF image, (C) Out-of-plane strain map indicating no gradual expansion of the o-o-p parameter as observed for uc-NNO2, (D) In-plane strain map.

5

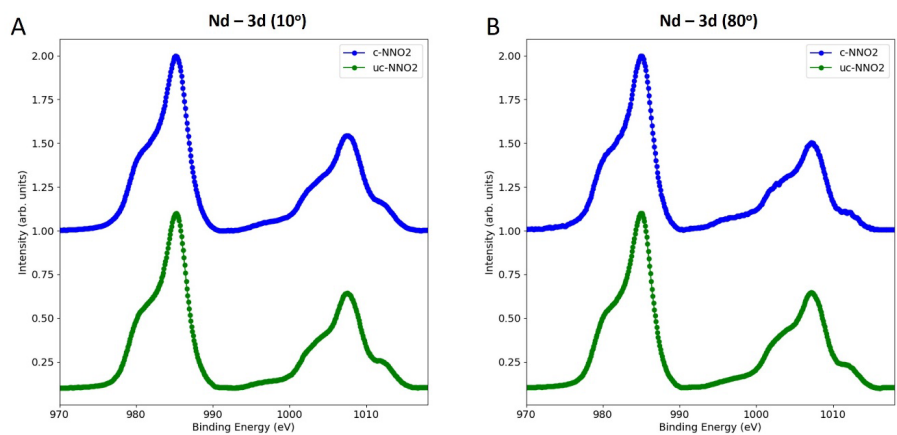

**Figure S5.** Hard X-ray Photoemission (HAXPES) comparison of the Nd 3d core level of the c-NNO2 and uc-NNO2 samples. (A) Comparison of Nd 3d photoemission in bulk sensitive 10º incidence condition and (B) surface sensitive 80º incidence condition from c-NNO2 and uc-NNO2 samples.

6


\end{document}